\documentclass{aa}  

\usepackage{graphicx}
\usepackage{color}
\usepackage{url}
\usepackage[colorlinks=true,citecolor=blue,linkcolor=blue]{hyperref}
\usepackage[normalem]{ulem}
\usepackage{txfonts}
\titlerunning{X-ray bursts of 4U 1636--536}
\usepackage{natbib}
\newcommand{\gtap}{\mathrel{\hbox{\rlap{\lower.55ex \hbox {$\sim$}}
                   \kern-.3em \raise.4ex \hbox{$>$}}}}
\newcommand{\ltap}{\mathrel{\hbox{\rlap{\lower.55ex \hbox {$\sim$}}
                   \kern-.3em \raise.4ex \hbox{$<$}}}}

\def\be{\begin{equation}} 
\def\ee{\end{equation}}

\begin{document}

\title{NICER observations of the evidence of Poynting-Robertson drag and disk reflection during type I X-ray bursts from 4U 1636--536}









\author{Guoying Zhao\inst{1}
        \and
        Zhaosheng Li\inst{1}
        \thanks{Corresponding author}
       \and
      Yuanyue Pan\inst{1}
             \and
    Maurizio Falanga \inst{2,3}
    \and
    Long Ji\inst{4}
    \and
    Yupeng Chen\inst{5}
    \and
    Shu Zhang\inst{5}
          }
   \offprints{Z. Li}

   \institute{Key Laboratory of Stars and Interstellar Medium, Xiangtan University, Xiangtan 411105, Hunan, P.R. China\\ \email{lizhaosheng@xtu.edu.cn}
               \and
International Space Science Institute (ISSI), Hallerstrasse 6, 3012 Bern, Switzerland
           \and
Physikalisches Institut, University of Bern, Gesellsschaftstrasse 6, 3012 Bern, Switzerland
              \and
School of Physics and Astronomy, Sun Yat-sen University, 519082 Zhuhai, Guangdong, China
              \and
Key Laboratory of Particle Astrophysics, Institute of High Energy Physics, Chinese Academy of Sciences, 19B Yuquan Road, Beijing 100049, China
              }
              
\date{Received 1 December 2021 / Accepted 14 February 2022}



 \abstract{
Type I X-ray bursts are the result of an unstable thermonuclear burning of accreting matter on the neutron star (NS) surface. The quick release of energetic X-ray photons during such bursts interacts with the surrounding accretion disk, which raises the accretion rate due to Poynting-Robertson drag and, thus, a fraction of the burst emission is reflected.  We analyzed two photospheric radius expansion bursts in the NS low-mass X-ray binary 4U 1636--536  that took place in 2017,  using data from Neutron star Interior Composition Explorer. The time-resolved burst spectra showed clear deviations from a blackbody model. The spectral fitting can be significantly improved by introducing either the enhanced persistent emission (the $f_a$ model)  or the reflection from the accretion disk (the \texttt{relxillNS} model). The  $f_a$ model  provides a higher blackbody temperature and higher burst flux compared with the \texttt{relxillNS} model. The peak fluxes of two bursts from the $f_a$ model, $4.36\times10^{-8}~\mathrm{erg~cm^{-2}~s^{-1}}$ and $9.10\times10^{-8}~\mathrm{erg~cm^{-2}~s^{-1}}$,  are slightly higher than the Eddington limits of mixed hydrogen-helium and pure helium bursts from previous observations, respectively. When the disk reflections have been taken into account simultaneously, the peak fluxes are lower to match the preferred values. We find evidence to support the finding that both the Poynting-Robertson drag and disk reflection have been appeared during these two X-ray bursts. Moreover, the disk reflection may contribute $\sim20-30\%$ of the total burst emissions. }


\maketitle

\section{Introduction\label{sec:intro}}
A neutron star (NS) in low-mass X-ray binary accretes matter from a Roche-lobe overflowing companion star with (sub-)solar mass. The accreted material is compressed and heated on the NS surface,  which occasionally triggers the unstable thermonuclear burning of helium, mixed hydrogen-helium, and carbon, also known as type I X-ray burst  \citep[superburst especially for carbon burning; see][for reviews]{Lewin93,Strohmayer06,Galloway21}. During an X-ray burst, the X-ray intensity first sharply increases, typically by a factor of 10 in  $\sim0.5-5$ s;  afterwards, it usually decreases exponentially within a few and up to $\sim10^3$ s \citep{Galloway20}. The total energy emitted from an X-ray burst is $\sim10^{39}-10^{41}$ erg. Some X-ray bursts are strong enough to lift up the NS surface layers, causing photospheric radius expansion (PRE), where the radiation flux that emerges from the NS surface reaches or slightly exceeds the Eddington limit \citep{Lewin93}. In the cooling phase of PRE bursts, the photosphere probably covers the whole NS surface, which provides a method for measuring the NS masses and radii, and to probe the equation of state of superdense matter \citep[see e.g.,][]{Sztajno87,Ozel09,Poutanen14,Li15,Li17,Suleimanov17b,Suleimanov20}.  




Careful spectral analyses have revealed that a large fraction of X-ray burst spectra are well fitted by a diluted blackbody model with a temperature of $T_{\rm bb}\sim 0.5-3 $ keV. The energetic X-ray photons released from such bursts interact with the surrounding accretion disk, the hot corona, and the NS atmosphere, bringing on several observational consequences \citep{Degenaar18,Fragile20}. The accretion rate can be increased during a burst due to the Poynting-Robertson drag, leading to enhanced persistent emission \citep{Walker92,Zand13,Worpel13}. Moreover, the bright X-ray burst irradiation can empty the material out of the inner accretion disk and, as a consequence, the accretion disk will take some time to restore itself, resulting in a temporary drop in persistent emission \citep{Bult21}. The emission lines and edges have been observed from the superbursts in 4U 1820--30 and IGR J17062--6143, which were alternatively explained by the reflection features from the photoionized accretion disk \citep[see, e.g.,][]{Ballantyne04, Keek17}.  The photoionization absorption edges were identified in 4U 0614+091, 4U 1722--30, and 4U 1820--30 \citep{Zand10}, as well as HETE J1900.1--2455 \citep{Kajava17} and GRS 1747--312 \citep{Li18} due to the bound--bound or bound--free transitions of the heavy elements 
generated in the burning ashes \citep{Weinberg06,Weisberg10}, allowing us to measure the gravitational redshift on the NS surface. In addition, frequent bursts are capable of cooling down hot corona in NS LMXBs, to accelerate the hard to soft spectral transition \citep{Li18b}.

The NS nature of 4U 1636--536 has been confirmed through the detection of type I X-ray bursts \citep{1977ApJ...217L..23H, galloway2006eddington, 2011MNRAS.413.1913Z}, as well as burst oscillations, indicating a spin frequency of $\sim581$ Hz \citep{1997IAUC.6541....1Z, giles2002burst, 2002ApJ...577..337S}. This source is classified as an atoll source \citep{hasinger1989two} with an orbital period of $\sim$ 3.8 h \citep{van1990orbital} and a companion star of mass $\sim0.4 M_\odot$, assuming an NS mass of $1.4M_\odot$ \citep{giles2002burst}. 

Since the discovery of 4U 1636--536, about 700 bursts (that number is rising), including four superbursts, have been observed, making it one of the best-studied X-ray bursters \citep[see e.g.,][]{Galloway20, Zand17}. Previous observations found two groups of PRE bursts from 4U 1636--536, which the bolometric flux peaked at $6.4\times10^{-8}~{\rm erg~cm^{-2}~s^{-1}}$ and $3.8\times10^{-8}~{\rm erg~cm^{-2}~s^{-1}}$ \citep{galloway2006eddington}, respectively. These authors proposed that the bright and faint PRE bursts were produced by the ignition of helium and mixed hydrogen-helium at approximately solar composition, respectively. The distance to 4U 1636--536 has been inferred at $6.0\pm0.5$ kpc  by assuming its brightest PRE bursts were powered by pure helium  \citep{galloway2006eddington,Galloway18}, which is consistent with the value of $4.42_{-1.63}^{+3.08}$ kpc  measured by {\it Gaia} \citep{Galloway20}. Thereafter, the distance of 6.0 kpc was adopted.   The enhanced persistent emissions during X-ray bursts have also been observed in 4U 1636--536 \citep{Worpel13,Worpel15,Roy21,Kashyap22}. However, the reflection features accompanying with the enhanced persistent emissions have only been detected during its superburst \citep{Keek14b,2018ApJ...855L...4K}.



In this paper, we report on the detailed analysis of the X-ray bursts from 4U 1636--536 using the data observed by the Neutron star Interior Composition Explorer (NICER). This work is organized as follows. We describe the general properties of the observed bursts in Sect. \ref{sec:floats}. The time-resolved spectral fitting results are provided in Sect. \ref{sec:spec}. We discuss the results in  Sect. \ref{sec:highlight} and present our conclusions in Sect.~\ref{sec:sum}.

\section{Observations} \label{sec:floats}

\begin{figure}[ht!]
\includegraphics[width=9.0cm]{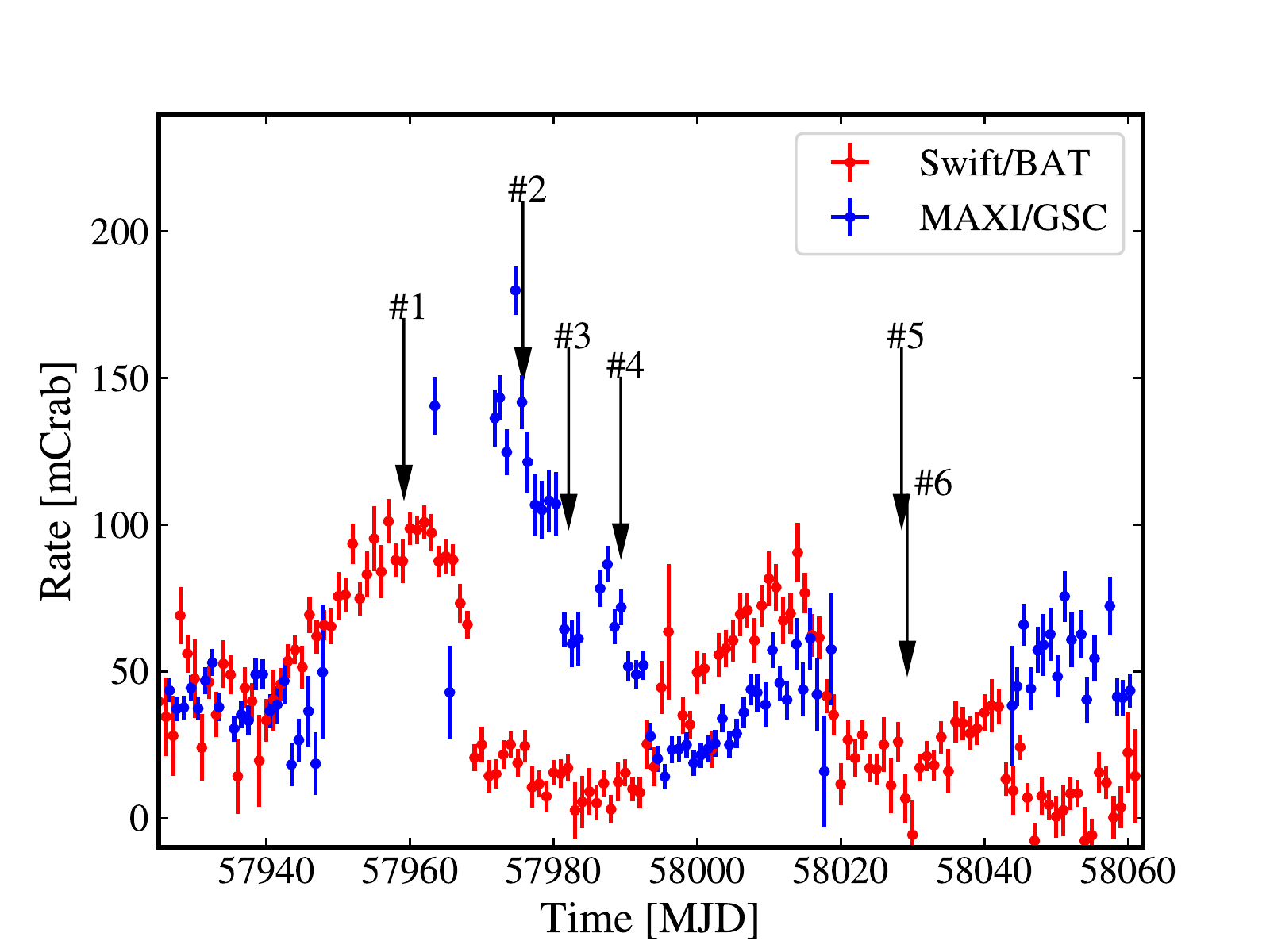}
\includegraphics[width=9.0cm]{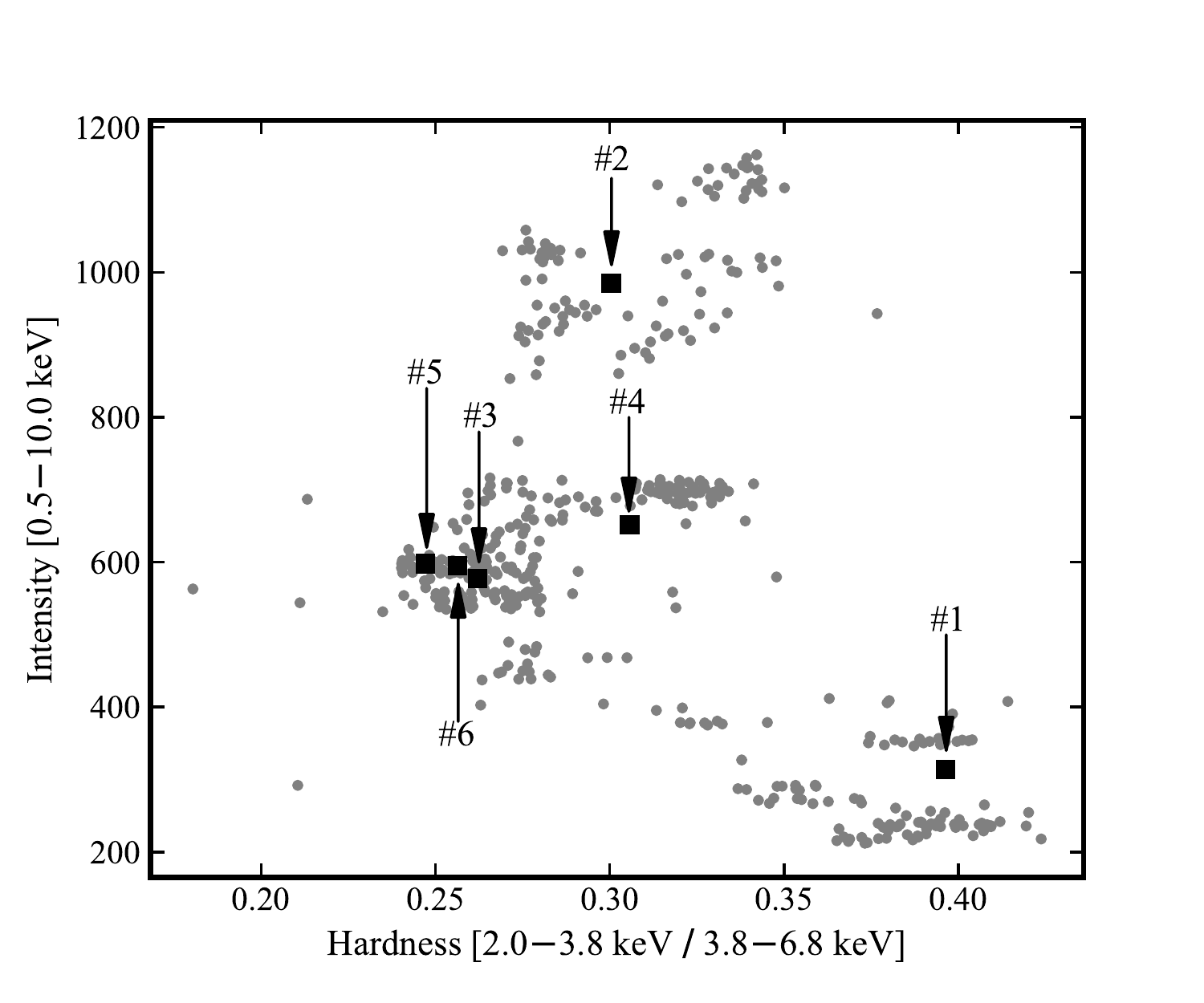}
\caption{Rate and spectral states of all six X-ray bursts. {\it Top panel}: {\it Swift}/BAT (red, 15--50 keV) and  MAXI/GSC (blue, 2--20 keV) light curves of 4U 1636--536 in units of Crab rate. The arrows represent six X-ray bursts observed by NICER. {\it Bottom panel}:\ HID from NICER observations. All bursts are removed, and each point represent a segment of 64 s. The HID of the persistent emission prior to each X-ray burst has been marked as black square. We focus on bursts \#4 and \#5.
\label{fig:outburst}}
\end{figure}

The {\it Neil Gehrels Swift} observatory satellite and the Gas Slit Camera \citep[GSC,][]{Mihara11} on board MAXI \citep{Matsuoka09} have been monitoring the X-ray activities of 4U 1636--536 in 2017. In Fig.~\ref{fig:outburst}, we show the light curves of 4U 1636--536 in 15--50 keV from {\it Swift}/BAT\footnote{\url{https://swift.gsfc.nasa.gov/results/transients/BAT_current.html}} and 2-20 keV from MAXI/GSC\footnote{\url{http://maxi.riken.jp/mxondem/}} in units of the Crab rate to illustrate the spectral state evolution. 

 NICER has the capabilities to detect X-ray photons in the energy of 0.2--12 keV with low-background, high-throughput, and high time resolution. It was launched and installed on the International Space Station on June 3, 2017 \citep{gendreau2017searching}. In this work, we analyze all 46 NICER observations of 4U 1636--536 on Modified Julian Date (MJD) 57926--58060, which have a total exposure time of 62.9 ks. We processed the NICER data using HEASOFT V6.29c and the NICER Data Analysis Software (NICERDAS). The default selection criteria were applied to filter the cleaned event data. We then applied barycentric coordinates to all the event data using the tool \texttt{barycorr} to the ICRS reference frame and JPL-DE405 ephemeris. We extracted  the light curves in the energies of 2.0--3.8 keV, 3.8--6.8 keV, and 0.5--10 keV, with a bin size of 1 s, and the spectra via the command \texttt{xselect}. We calculated the hardness between 2.0--3.8 keV and 3.8--6.8 keV, and we constructed the intensity-hardness diagram (HID) with the X-ray bursts removed (see Fig.~\ref{fig:outburst}).

\begin{figure}[ht!]
\includegraphics[width=9.0cm]{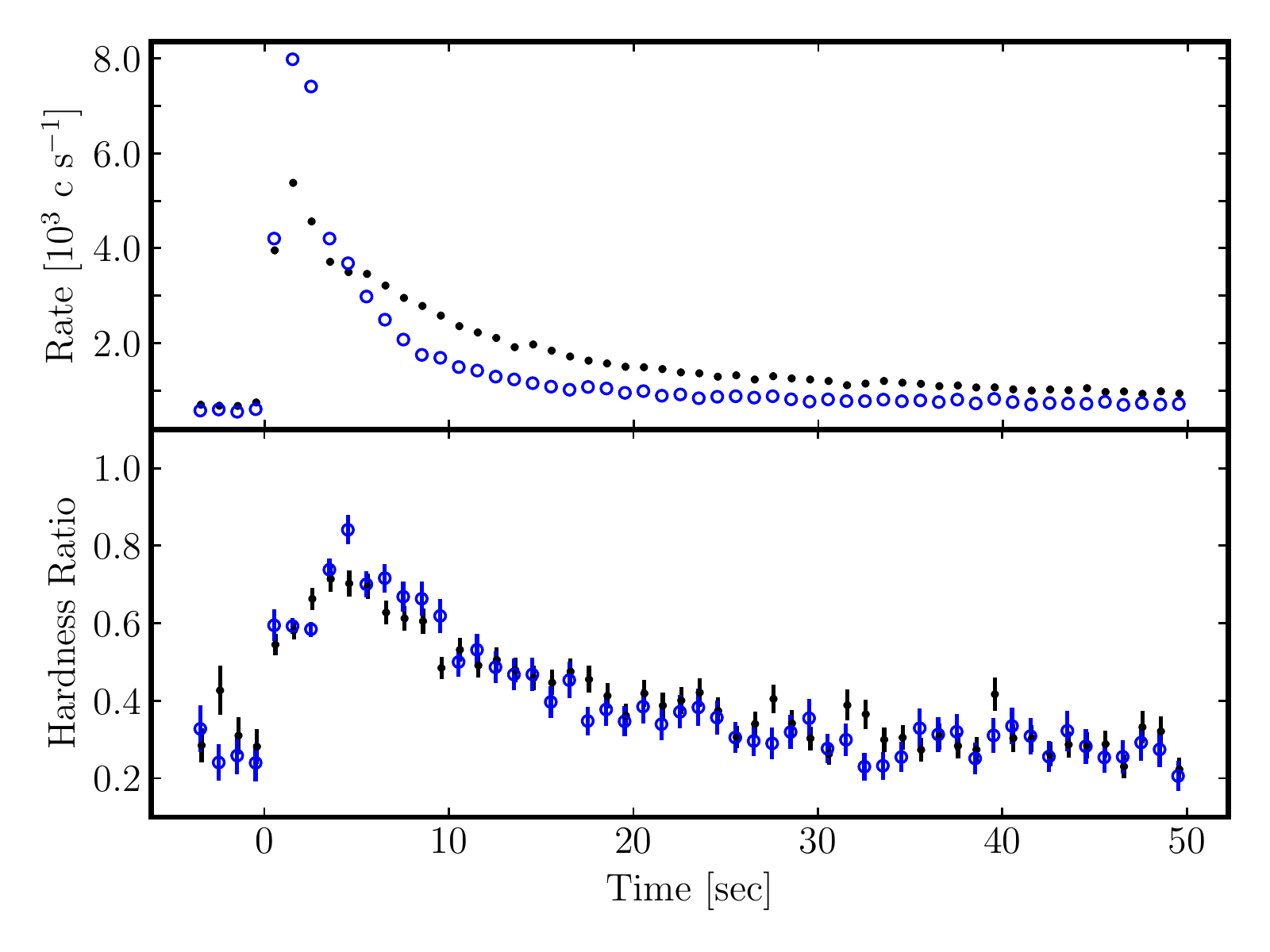}
\caption{Light curves and hardness ratio of bursts \#4 and \#5. {\it Top panel}: 1 s light curves of bursts \#4 (black dots) and \#5 (blue circle) in the 0.5--10 keV, triggered on MJD 57989.3665 and 58028.4772, respectively. {\it Bottom panel}: hardness ratio between 2.0--3.8 keV and 3.8--6.8 keV.
\label{fig:lc}}
\end{figure}

In total, we identified six type I X-ray bursts  identified,\footnote{We note that the cleaned event file produced from \texttt{nicerl2} only kept the decay of the X-ray burst in ObsID 1050080103. However, the whole X-ray burst is clearly shown in the unfiltered event file.} where the onset of all bursts are marked in Fig.~\ref{fig:outburst}. The first X-ray burst occurred around the peak of the {\it Swift}/BAT light curve and greater hardness with lower intensity, indicating the hard spectral state of the persistent emission, whereas another five X-ray bursts triggered in soft spectral states, since the {\it Swift}/BAT rates were rather low and the hardness were minimal. In this work, we focus on two brightest type I X-ray bursts, namely: burst \#4 in ObsID 1050080126 at MJD 57989.3665 (2017 August 24), and burst \#5 in ObsID 1050080128 at MJD 58028.4772 (2017 October 2), which have peak rates of $5.3\times10^{3}~\rm{c~s}^{-1}$ and $8.1\times10^{3}~\rm{c~s}^{-1}$ in 0.5--10 keV, respectively. The light curves and the hardness ratio between 2.0-3.8 keV and 3.8--6.8 keV of these two bursts are shown in Fig. \ref{fig:lc}. The hardness ratios are around 0.25 before the bursts, increasing to $\sim$ 0.9 in $\sim$ 6 s and then evolving to the pre-burst level at end of the bursts. The maximum of the hardness ratio occurred $\sim$ 4 s later than the peak of the light curves, which implies the spectral evolution during bursts, see Sect.~\ref{sec:spec}. We searched for burst oscillations in the frequency range 576--586 Hz by applying the $Z_1^2$-test statistics based on {\tt Stingray} \citep{Buccheri83, Huppenkothen19}. The cleaned event files in the 0.5--2 keV, 2--10 keV, and 0.5--10 keV energy ranges, respectively, are used by applying a moving window method with the window sizes $T$ of 2 s, and steps of $T/10$.  We did not find any signal for the burst oscillation around 581 Hz. 


We performed the spectral analysis using Xspec 12.12.0 \citep{1996ASPC..101...17A}. The 0.3--10 keV spectra are grouped with a minimum of 20 counts per channel using the ftools task \texttt{grppha}. The background spectra are generated via the \texttt{nibackgen3C50} tool \citep{Remillard21}. The ancillary response files (ARFs) and response matrix files (RMFs) are generated from the tools \texttt{nicerarf} and \texttt{nicerrmf}, respectively. The errors of all parameters are quoted at the $1\sigma$ confidence level.

\section{Results} \label{sec:spec}
\subsection{Persistent emission} \label{subsec:floats}

For two bursts in 4U 1636--536, we extracted the persistent spectra in the energy range of 0.3--10 keV, with an exposure time of 64 s prior to the trigger, which includes photons from the source and the  background from the instrument and the sky.  We first fit the persistent spectra with a simple and phenomenological cutoff power-law model, \texttt{cutoffpl}, and then with a thermally Comptonized continuum model, \texttt{nthcomp} \citep{Zdziarski96,Zycki99}, both modified by the Tübingen–Boulder model, \texttt{TBabs}, with abundances from \cite{Wilms_2000}. The latter model has been usually used to fit the continuum spectra of 4U 1636--536 in both hard and soft spectral states \citep[see e.g.,][]{Lyu14,Wang17}. The parameters are the equivalent hydrogen column, $N_{\rm H}$, for \texttt{TBabs}; the photon index and e-folding energy of exponential rolloff, $\Gamma$ and $E_{\rm cut}$, for \texttt{cutoffpl};  the asymptotic power-law photon index, $\Gamma$, the electron temperature, $kT_{\rm e}$, the seed photon temperature, $kT_{\rm BB}$, the type of seed photons (0 or 1 for blackbody or disk-blackbody seed photons, respectively), and the normalization, for \texttt{nthcomp}.

\begin{table}
\caption{\label{tab:Fpers}Best-fitting parameters for the persistent spectra prior to bursts \#4 and \#5, by using \texttt{TBabs$\times$cutoffpl} and \texttt{TBabs$\times$nthcomp}.}
\begin{tabular}{cll}
\hline 
Obs Id & 1050080126  & 1050080128 \tabularnewline
\hline

\hline 
&\texttt{TBabs$\times$cutoffpl}   & \tabularnewline

\hline 
$N_{\mathrm{H}}\,(10^{22}~\mathrm{cm^{-2}})$
& $0.50\pm0.01$ & $0.52\pm0.01$ \tabularnewline
$\Gamma$ & $1.45\pm0.07$ & $1.69\pm0.07$\tabularnewline
$E_{\mathrm{cut}}\,(\mathrm{keV})$  & $8.19_{-1.2}^{+1.6}$  &$6.35_{-0.92}^{+1.27}$ \tabularnewline

$\chi_{\rm red}^{2}\,(\mathrm{degrees\, of\, freedom})$  & $1.07\,(471)$ & $1.00\,(418)$\tabularnewline

\hline 
&\texttt{TBabs$\times$nthcomp}   & \tabularnewline
\hline 
$N_{\mathrm{H}}\,(10^{22}~\mathrm{cm^{-2}})$
& $0.41\pm0.01$ & $0.41\pm0.01$ \tabularnewline
$\Gamma$ & $1.96\pm0.03$ & $2.34\pm0.05$\tabularnewline
$kT_{\mathrm{e}}\,(\mathrm{keV})$$^{\mathrm{a}}$  & $5.6_{-1.7}^{+p}$  &$13.6_{-6.3}^{+p}$ \tabularnewline
$kT_{\mathrm{BB}}\,(\mathrm{keV})$  & $0.52\pm0.03$ & $0.47\pm0.02$\tabularnewline
$\chi_{\rm red}^{2}\,(\mathrm{degrees\, of\, freedom})$  & $1.05\,(470)$ & $0.92\,(417)$\tabularnewline
\hline
$F_{0.3-10}\,(\mathrm{10^{-9}~erg\, s^{-1}\, cm^{-2}})$  & $2.85\pm0.04$ & $2.31\pm0.02$\tabularnewline

\hline

\hline

\end{tabular}
\label{table:burst}
\tablefoot{$^{\mathrm{(a)}}$ The symbol $p$ means that the poorly constrained electron temperature is pegged at hard limit.}
\end{table}

\begin{figure}[ht!]
\includegraphics[width=9cm]{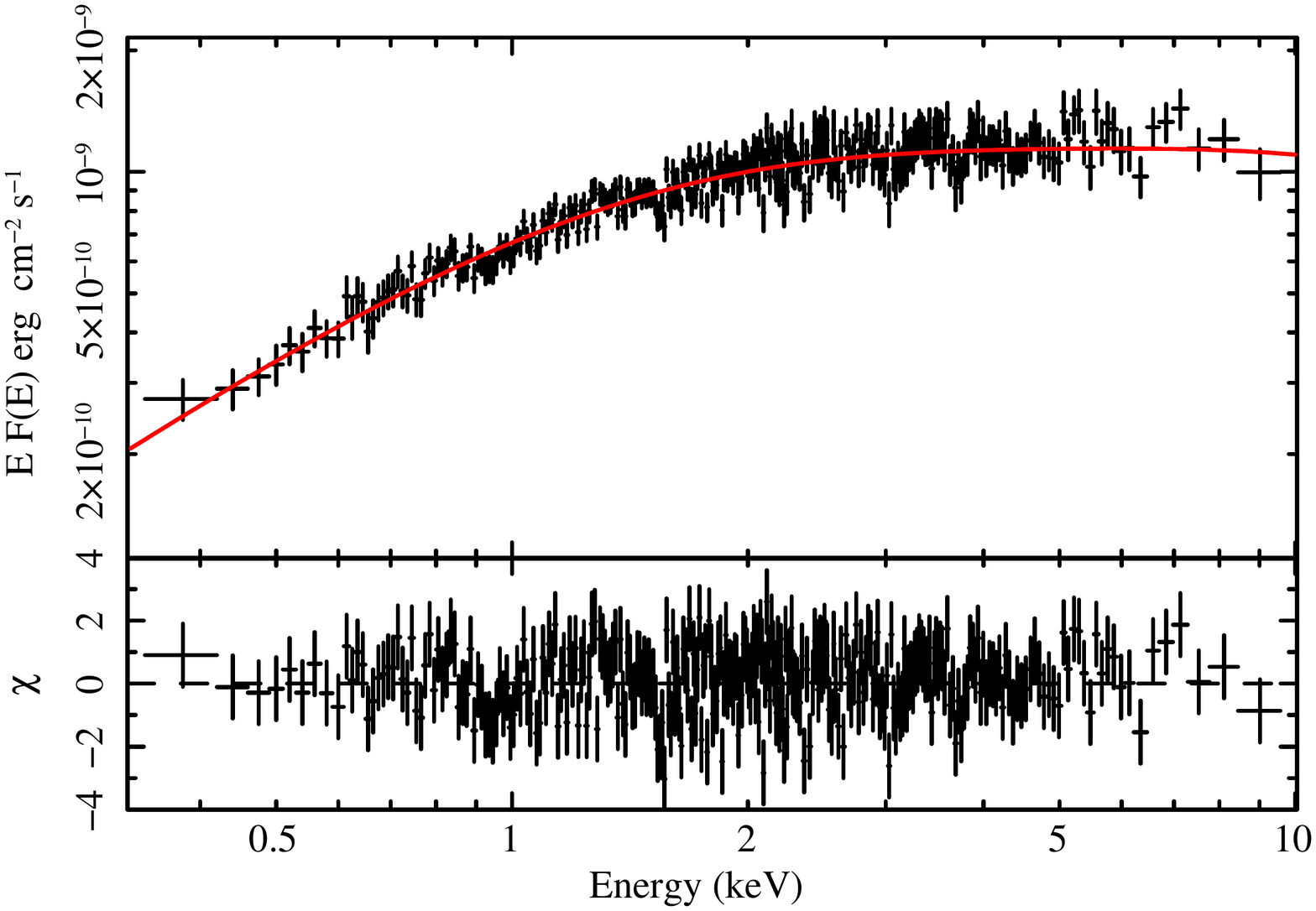}

\includegraphics[width=9cm]{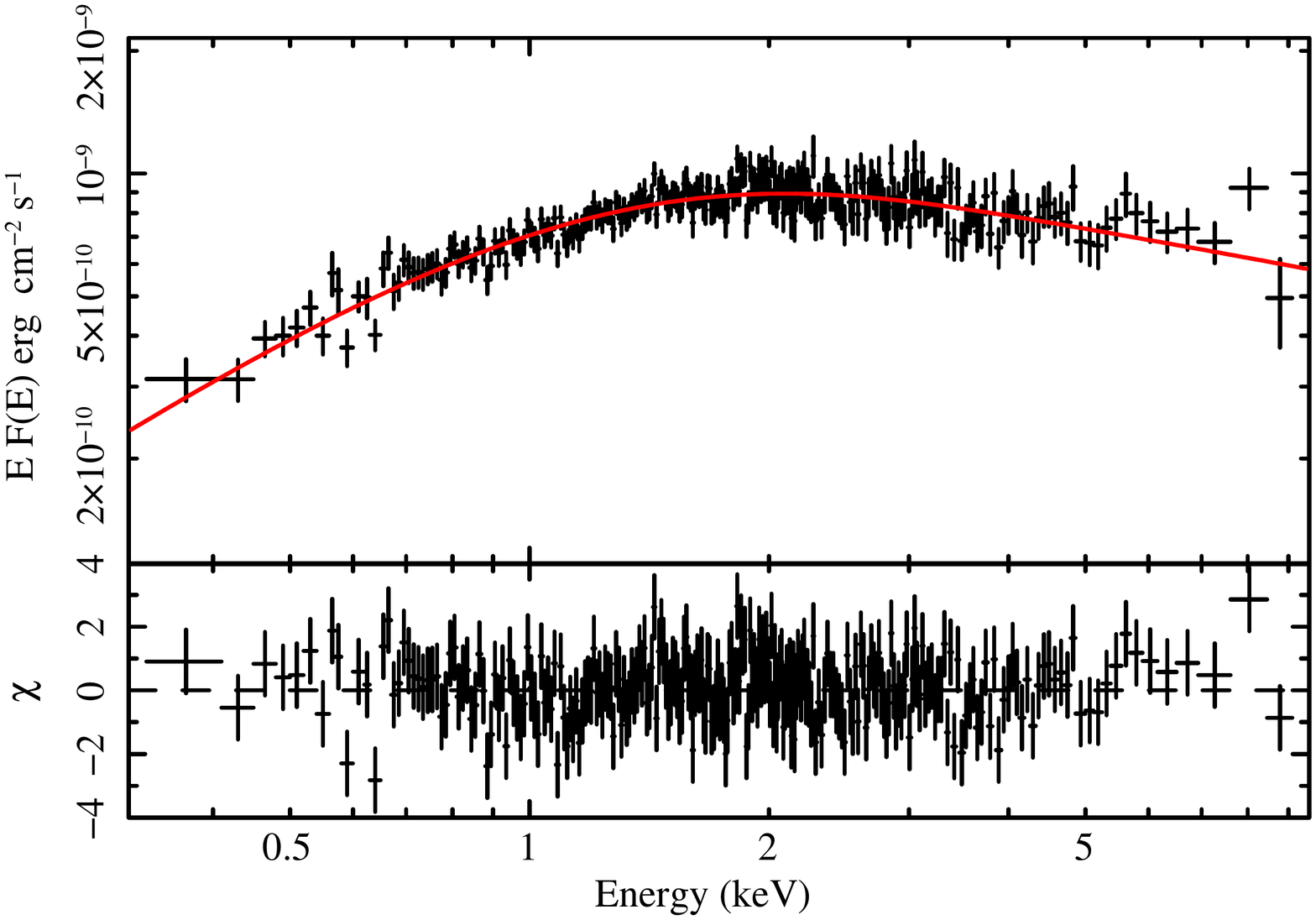}
\caption{0.3--10 keV persistent spectra and their best fitted models, {\tt TBabs$\times$nthcomp}, for bursts \#4 ({\it top panel}) and \#5 ({\it bottom panel}). The residuals related to the best fitted models for each spectrum are also shown.
\label{fig:persistent}} 
\end{figure} 

The optimally determined parameters are listed in Table~\ref{tab:Fpers} for the two selected models. Both the \texttt{cutoffpl} model and the \texttt{nthcomp} model can fit two persistent spectra well for the chi-square per degree of freedom (d.o.f.), $\chi^2_{\rm red}$, which is close to 1.0. We prefer the results from the \texttt{nthcomp} model, which is physically motivated. In Fig.~\ref{fig:persistent}, we show the best fit models to the persistent spectra of ObsID 1050080126 (top panel) and ObsID 1050080128 (bottom panel). We assumed the type for the disk's blackbody seed photons for the \texttt{nthcomp} model. The electron temperature cannot be constrained tightly due to a lack of data above 10 keV. Moreover, all parameters are remotely comparable with the previous results \citep[see e.g.,][]{Wang17,Mondal21}. We note that two persistent emissions showed a quite similar spectral shape. The hydrogen column densities, $(0.41\pm0.01)\times10^{22}~\mathrm{cm^{-2}}$, are consistent with the result from \citet{Willingale13},\footnote{\url{https://www.swift.ac.uk/analysis/nhtot/}} but larger than the value, $0.25\times10^{22}~\mathrm{cm^{-2}}$, used in \citet{Galloway08}. We also calculated the unabsorbed flux in the energy ranges of 0.3--10 keV  by using the tool \texttt{cflux}. The bolometric flux is $(3.63\pm0.06)\times \mathrm{10^{-9}~erg\, s^{-1}\, cm^{-2}}$  corresponding to  $\sim4\%$  of the Eddington limit (the empirical critical luminosity,  $L_{\rm Edd}=3.8\times10^{38}~\rm{erg~s^{-1}}$, adopted from \citealt{Kuulkers03}), respectively. 


\subsection{X-ray burst time-resolved spectroscopy} \label{subsec:tables}

\begin{figure}
\includegraphics[width=9cm]{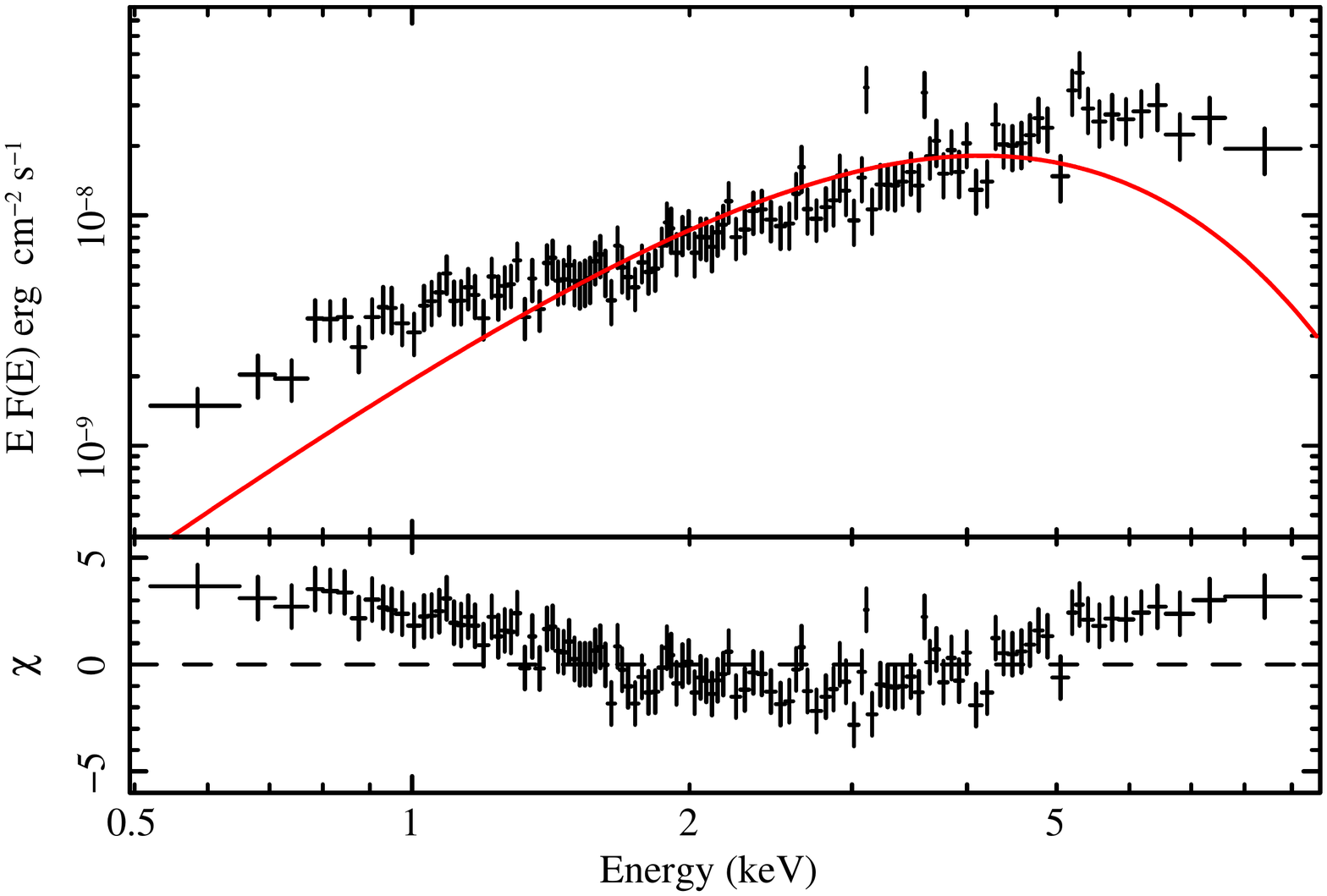}

\includegraphics[width=9cm]{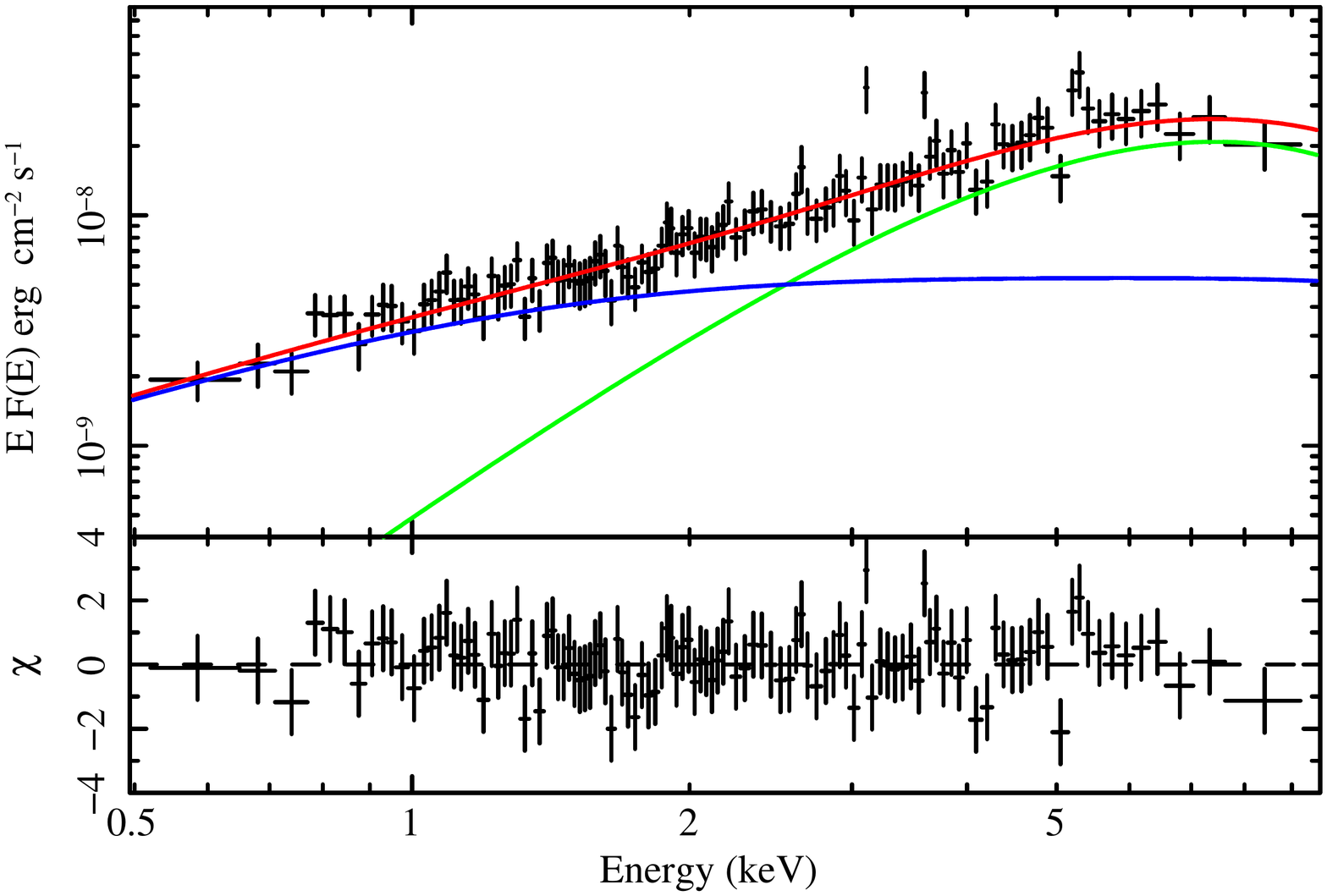}

\includegraphics[width=9cm]{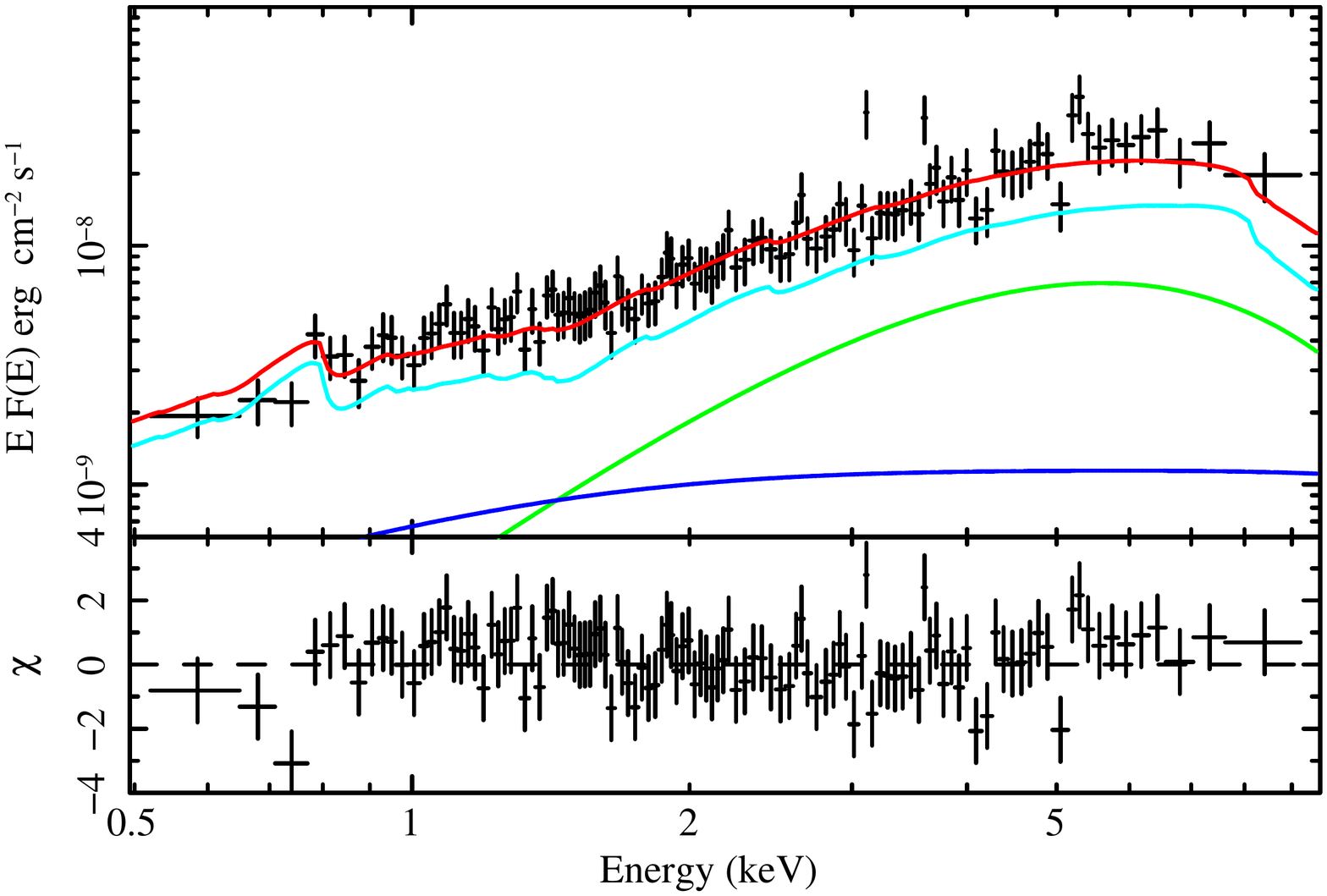}

\caption{Spectrum near the peak of burst \#4 and the best fitting models. \textit{Top panel}: spectrum is fitted with the blackbody model. \textit{Middle panel}: Increasing the normalization of the pre-burst emission can fit the spectrum well. \textit{Bottom panel}: additional disk reflection model can also produce a similar improvement. The red, green, blue, and cyan curves represent the total model, the \texttt{nthcomp}, the blackbody, and the reflection component, respectively. 
\label{fig:bb_burst1}} 
\end{figure}

\begin{figure}
\includegraphics[width=9cm]{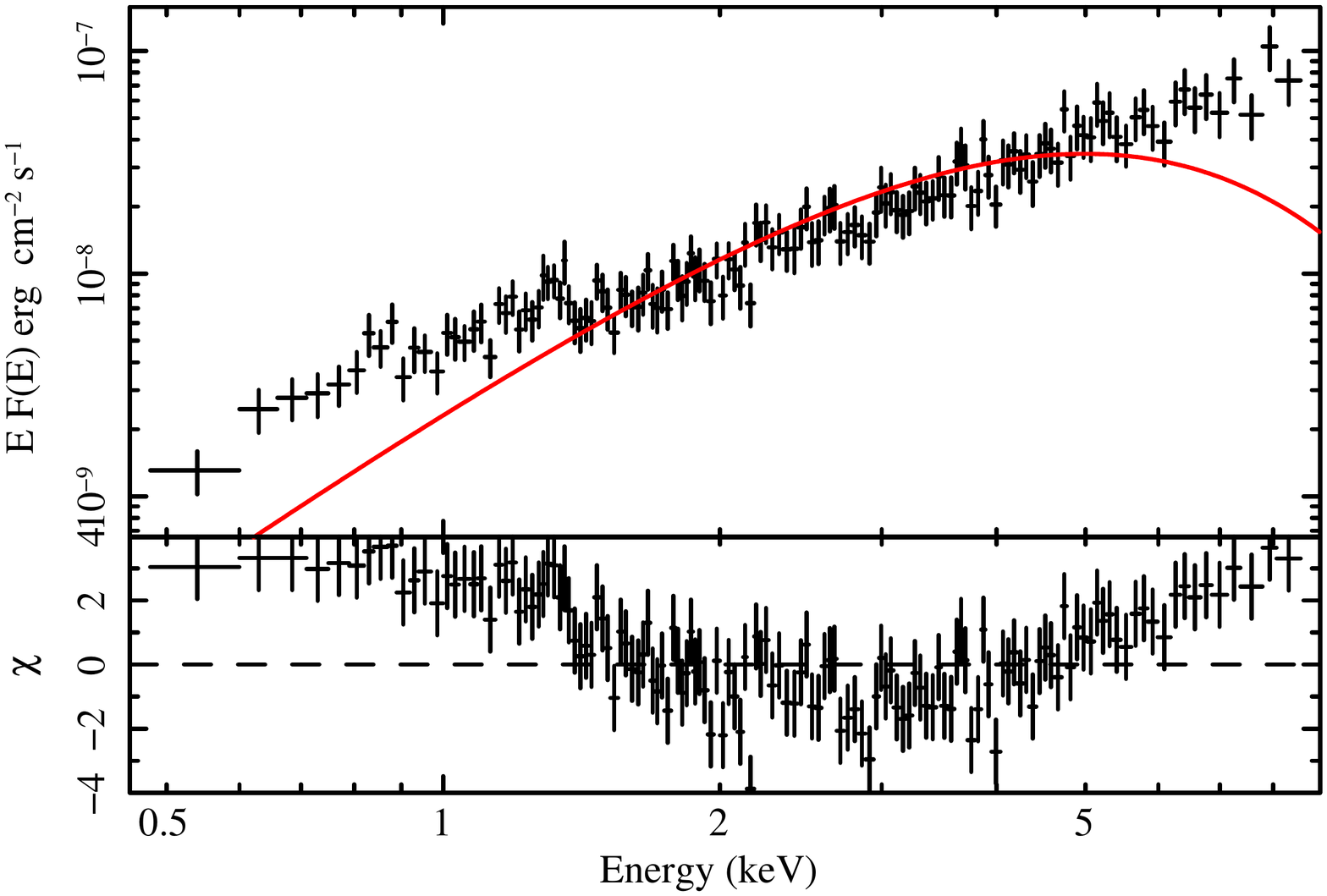}

\includegraphics[width=9cm]{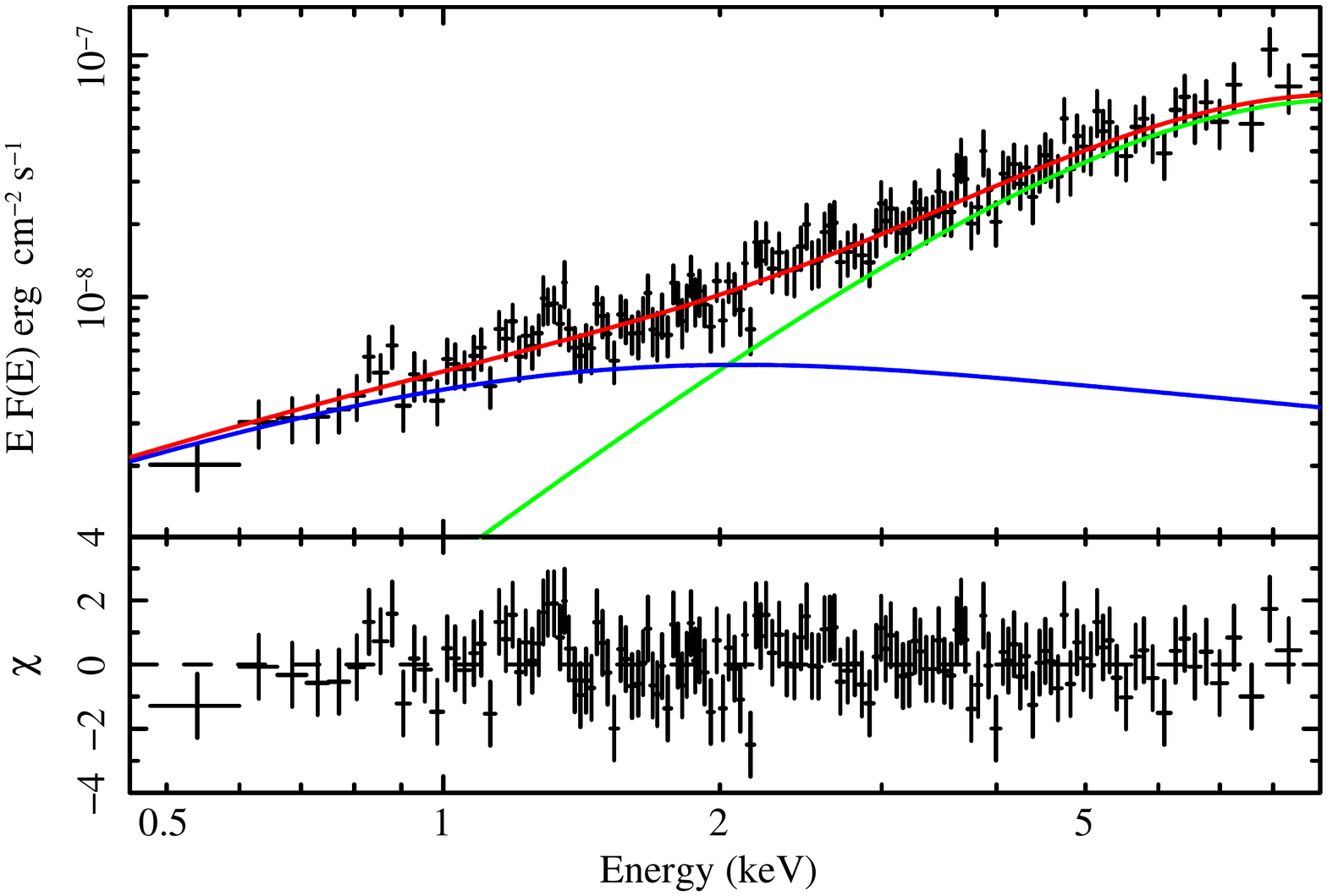}

\includegraphics[width=9cm]{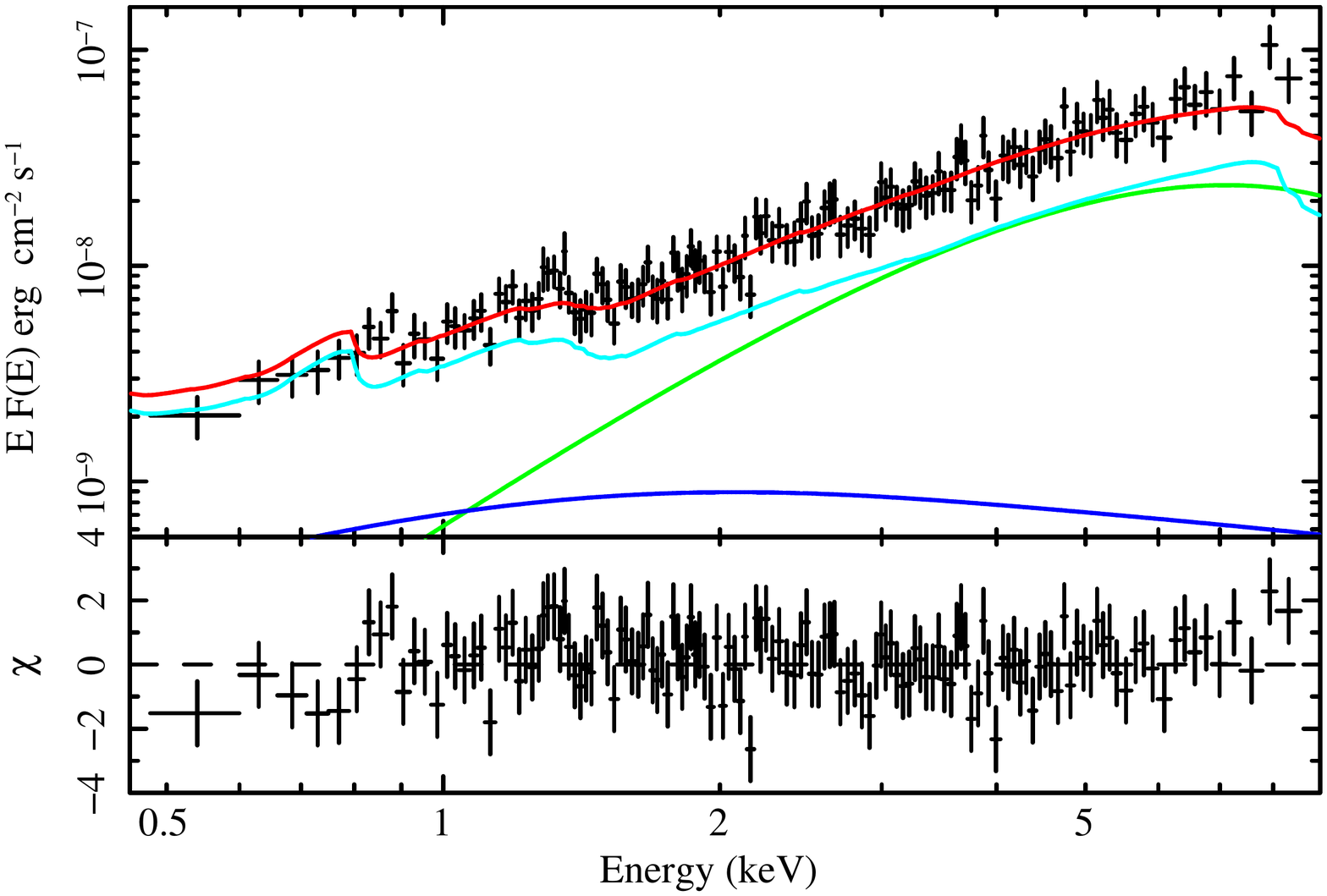}
\caption{Spectrum near the peak of burst \#5 and the best fitting models.
Details are similar to the description of Fig.~\ref{fig:bb_burst1}. 
\label{fig:bb_burst2}} 
\end{figure}

To investigate the spectral evolution during the bursts, the time-resolved spectra in the energy range of 0.3--10 keV are extracted with varying exposure time of 0.5--4 s, to make sure a total count of 2000 or more in each spectrum. We first adopted an absorbed blackbody model, \texttt{TBabs*bbodyrad}, to fit the burst spectra, where the persistent spectra are regarded as background. The \texttt{bbodyrad} model has two parameters: the blackbody temperature, $T_{\rm bb}$, and normalization, $K_{\rm bb} = (R_{\rm bb}/D_{10})^2$, where $R_{\rm bb}$ is the blackbody radius in units of km and $D_{\rm 10}$ is the distance to the source in units of 10 kpc. We fixed the hydrogen column density at $0.41\times10^{22}~\mathrm{cm^{-2}}$. We took one burst spectrum around the peak from each burst as examples and then fit these burst spectra with the blackbody model (see the top panels in Figs.~\ref{fig:bb_burst1} and \ref{fig:bb_burst2}). The residuals show strong  deviations below 1.5 keV and above 5 keV. Moreover, the model cannot fit the spectra well during the first $\sim 10$ s for these two bursts (see the grey points in Fig.~\ref{fig:fa}). The fitting results can be significantly improved by introducing enhanced persistent emission during bursts (see Sect.~\ref{sec:fa}) or an additional component reflected from the surrounding accretion disk (see Sect.~\ref{sec:reflection}).


\subsubsection{Enhanced persistent emission, $f_a$}
\label{sec:fa}
To investigate the enhanced persistent emission due to the Poynting–Robertson drag, the modeled instrumental background was subtracted from the burst emission. Then we adopted the $f_a$ model, \texttt{TBabs$\times$(bbodyrad+${f_a}\times$nthcomp)}, to fit the burst spectra, where the multiplication factor ${f_a}$ accounts for the increased persistent emission and the parameters of \texttt{nthcomp} are fixed to the best-fit values as listed in Table~\ref{tab:Fpers}. The parameter ${f_a}$ illustrates the contribution of the persistent emission, that is, ${f_a}=1$ indicates that the amplitude of the persistent emission is exactly the same as the moment before the X-ray burst. We note that this method is equivalent to only set the normalization of \texttt{nthcomp} free. The goodness of fit are $\chi_{\mathrm{red}}^2/\mathrm{d.o.f.}=0.82/112$ and $\chi_{\mathrm{red}}^2/\mathrm{d.o.f.}=0.86/137$, and  ${f\rm_a}=4.69\pm0.25,$ and ${f\rm_a}=6.52\pm0.42$, for the burst spectra around the peak of bursts \#4 and \#5, respectively (see the middle panels of Figs.~\ref{fig:bb_burst1} and \ref{fig:bb_burst2}). Compared with the results from the blackbody model, we have very small F-statistic probabilities (less than $10^{-19}$), which means the ${f_a}$ model provides significant improvements.   



\begin{figure*}[ht!]
\includegraphics[width=9.cm]{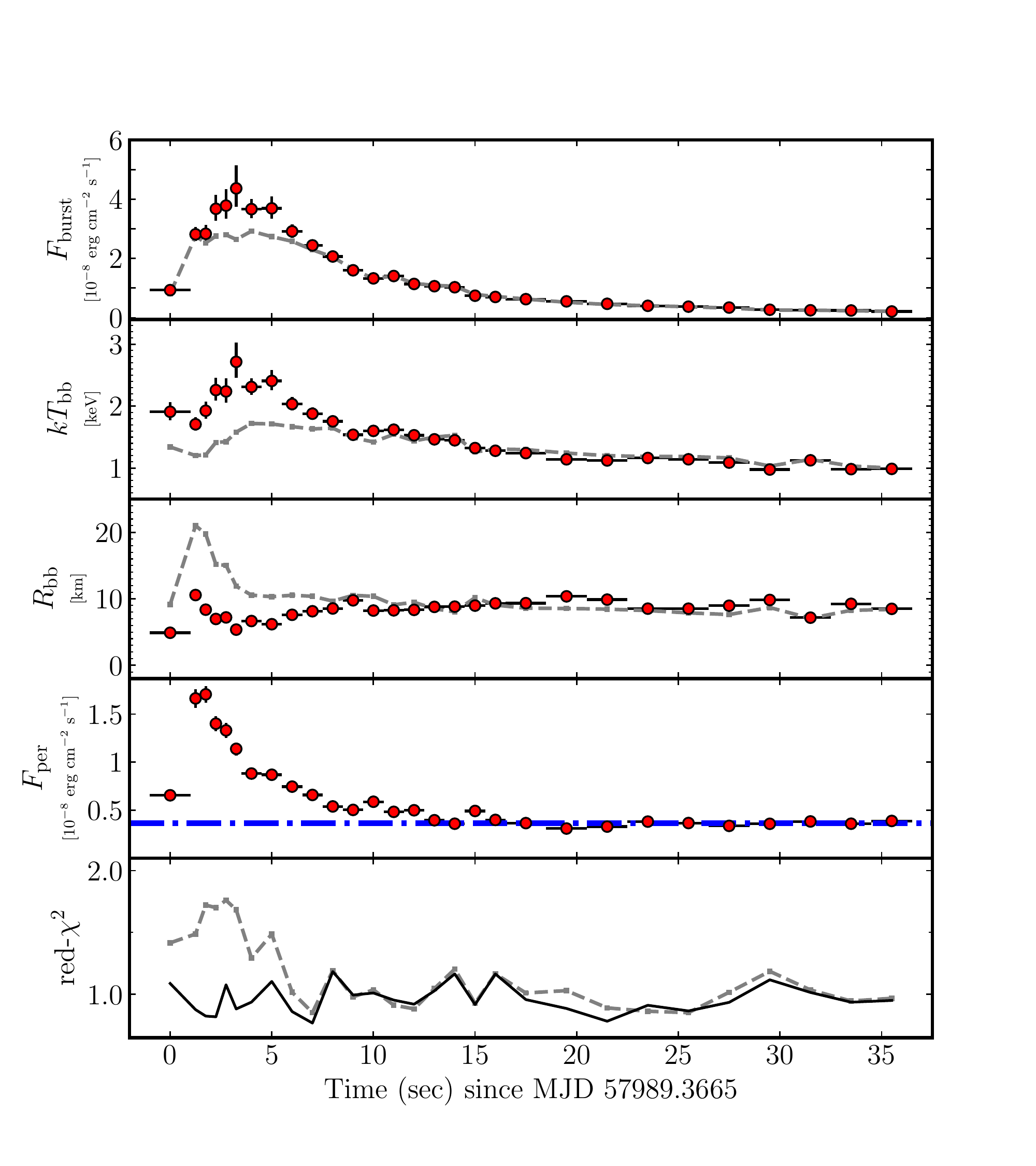}
\includegraphics[width=9.cm]{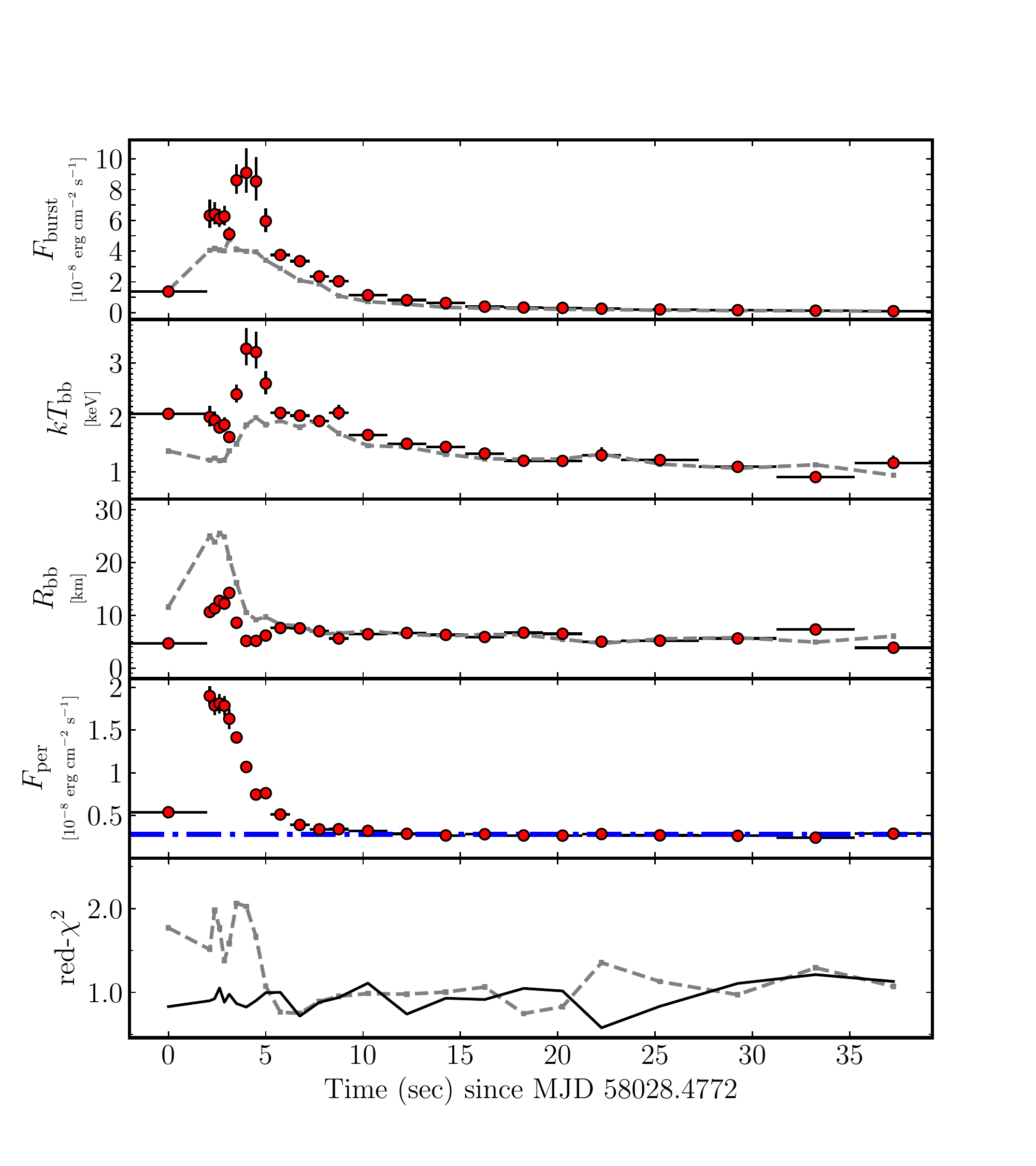}
\caption{Best-fit values from the $f_a$ model, \texttt{TBabs$\times$(bbodyrad + ${f_a}\times$nthcomp)}, and 1$\sigma$ uncertainties of two X-ray burst spectra. {\it From top to bottom}:\ burst flux, $F_{\rm burst}$, the blackbody temperature, $kT\rm_{bb}$, and radius, $R\rm_{bb}$, the enhanced persistent emission flux, $F_{\rm per}$, and the goodness of fit per degree of freedom, ${\chi_{\rm red}^2}$, for bursts \#4 \textit{(Left Panel)} and \#5 \textit{(Right Panel)}. The blue dashed-dot lines represent the flux of the persistent emissions before the bursts. The grey data are fitted by the blackbody model for comparisons.
\label{fig:fa}}
\end{figure*}

In Fig.~\ref{fig:fa}, we show the best fit of all the parameters with the $f_a$ model, and the results of the blackbody model are plotted as grey dots as a comparison. The $\chi^2_{\rm red}$ are around 1.0, which suggests that all the burst spectra fitted well with the ${f_a}$ model.  At the beginning of the two bursts, the ${f_a}$ value rises rapidly and reaches to their maximum values, 4.7 and 6.8, respectively, then decreases slowly and returns to around 1 during the tail. This indicates that the persistent emissions were rapidly increased in response to X-ray bursts during the expansion phase, and returned to the pre-burst level during the cooling tails.  For the bursts \#4 and \#5, the bolometric flux peaks at $(4.36\pm0.62)\times10^{-8}~\mathrm{erg~cm^{-2}~s^{-1}}$ occurred $\simeq3$ s after its onset, and exponentially decay with the time of $\tau=5.82\pm0.29$ s.  Meanwhile, the blackbody temperature and radius reached their maximum and minimum, respectively. 
The burst \#5 has higher peak bolometric flux and shorter decay time, namely, $(9.10\pm1.31)\times10^{-8}~\mathrm{erg~cm^{-2}~s^{-1}}$ and $2.44\pm0.20$ s, respectively. Then we obtain the burst fluences from multiplying the peak flux by the {\it e}-folding time, see Table~\ref{table:burst}. The blackbody radius and temperature reached the maximum  and minimum simultaneously for these two bursts at first few seconds. And then, the blackbody radius contracted to the minimum, 5.4 km and 5.1 km,  meanwhile, the blackbody temperatures increased to the maximum at $\sim 2.7$ keV and $\sim 3.3$ keV, respectively.  Based on the criteria in \citet{Galloway08}, they belongs to PRE bursts. During the cooling tail, the blackbody radii remain unchange around 8.9 km and 6.8 km. The increasing of radius from the peak of the burst to the cooling tail were caused by the variation of the color correction factor \citep[][]{Suleimanov12}.

\begin{table*}
\centering
\small
\caption{Properties of two PRE bursts. 
}
\begin{tabular}{lccccccc} 
\hline 
Burst ObsID  & Start Time  &  Rise Time  & Peak Rate& Peak Flux & Persistent Flux & $\tau$ & Burst Fluence \\
&   MJD & s & $\rm{c~s^{-1}}$ & $10^{-8}~{\rm erg~cm^{-2}~s^{-1}}$ & $10^{-9}~{\rm erg~cm^{-2}~s^{-1}}$& s &   $10^{-7}~{\rm erg~cm^{-2}}$\\ 
\hline 
\noalign{\smallskip}  
1050080126  & 57989.3665  &  $\sim3$  & $5.3\times10^3$ & $4.36\pm0.62$ & $3.63\pm0.06$ &  $5.82\pm0.29$  & $2.54\pm0.38$ \\
1050080128  &58028.4772  &  $\sim3$  &$8.1\times10^3$  & $9.10\pm1.31$  & $2.78\pm0.06$ &  $2.44\pm0.20$   &$2.22\pm0.37$ \\
\noalign{\smallskip}  
\hline  
\end{tabular}  
\label{table:burst}
\tablefoot{The peak flux, persitent flux, $\tau$, and burst fluence are determined from the $f_a$ model.}
\end{table*} 



%




\subsubsection{Disk reflection}
\label{sec:reflection}

\begin{figure*}[ht!]
\includegraphics[width=9.cm]{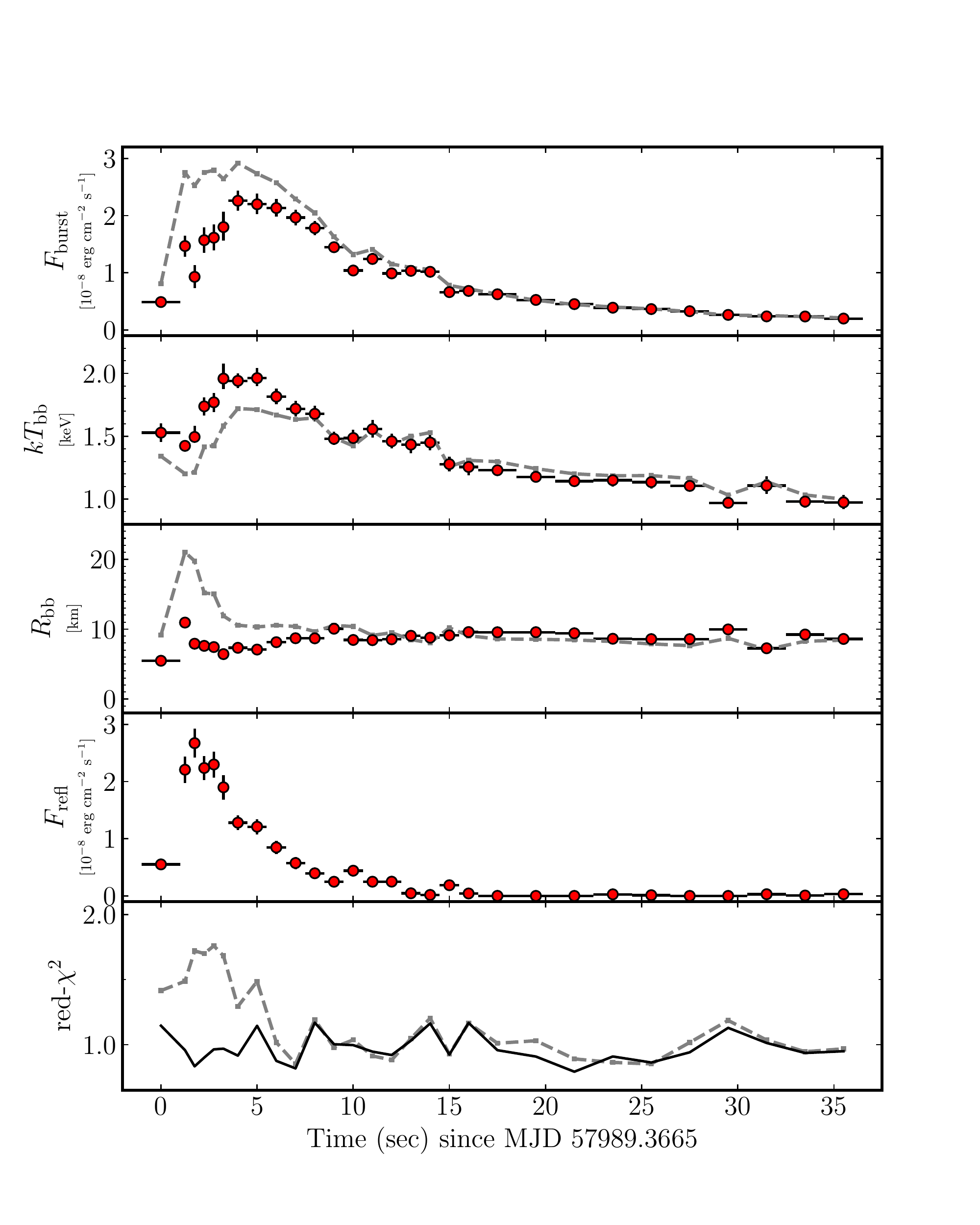}
\includegraphics[width=9.cm]{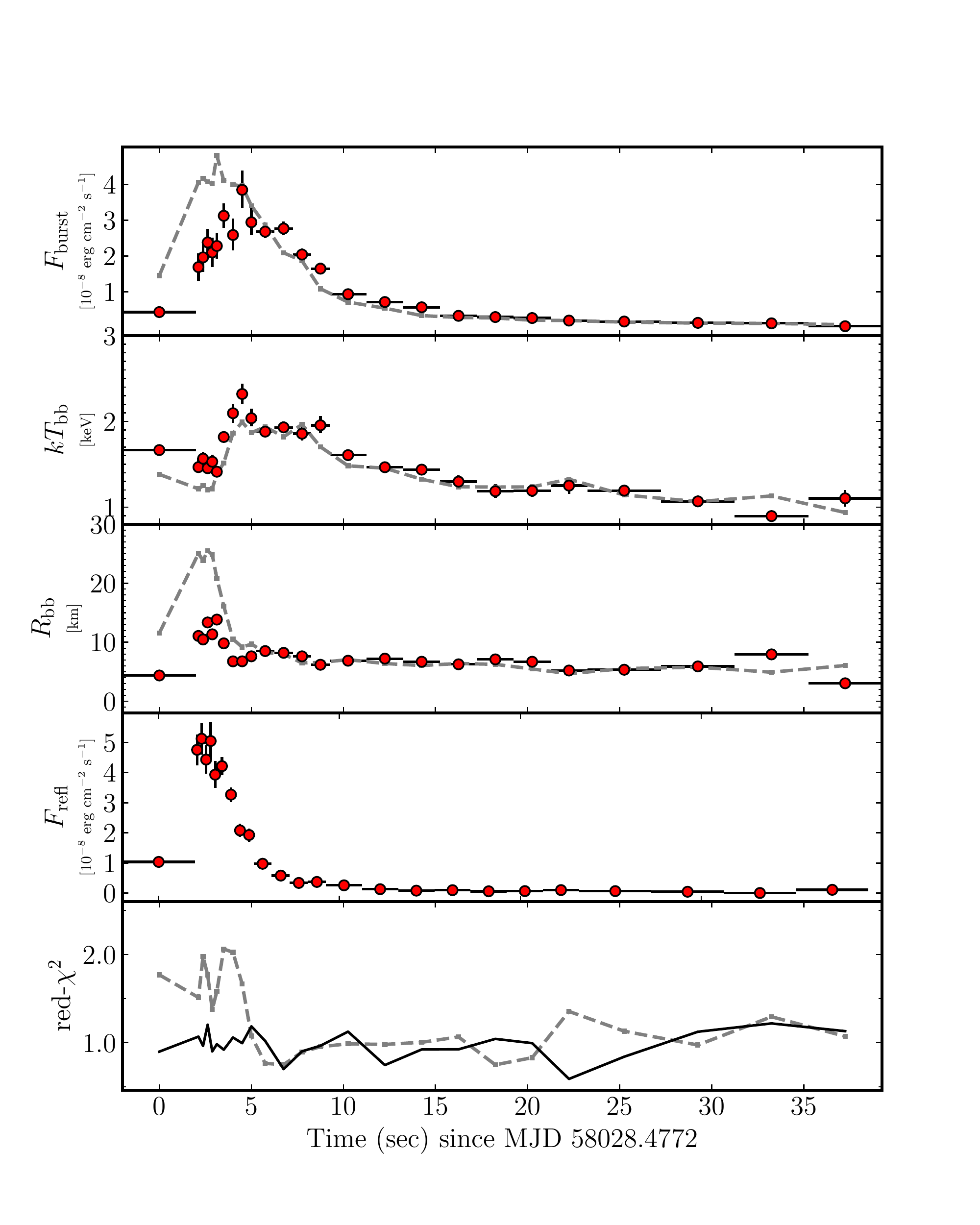}
\caption{ Results from the reflection model  \texttt{TBabs$\times$(bbodyrad + relxillNS + nthcomp)}.
Details are similar to the description of Fig. 6. \label{fig:refl}}
\end{figure*}

The deviations from the blackbody model of the burst spectra can also be explained by the disk reflection. In order to correctly characterize the reflection features in the burst spectra, we used the latest relativistic reflection model, \texttt{relxillNS},\footnote{\url{http://www.sternwarte.uni-erlangen.de/~dauser/research/relxill/}} of a photoionized accretion disk illuminated by a blackbody spectrum from NS \citep{Garcia21}. This model has been successfully fitted to the continuum of NS and black hole LMXBs \citep{Ludlam19,Connors20}. Here, for the first time, we apply the \texttt{relxillNS} model to account for the contribution from the accretion disk reflecting the burst radiation. The complete Xspec model is \texttt{TBabs$\times$(bbodyrad + relxillNS + nthcomp)}, where \texttt{bbodyrad},  \texttt{relxillNS,} and \texttt{nthcomp} represent the components from the burst, the reflection of the burst photons from the accretion disk, and the persistent emission, respectively. In this case (same as the procedure described in Sect.~\ref{sec:fa}), only the instrument background is subtracted in the fitting. The parameters of the \texttt{nthcomp} component is fixed to the best fitted values.  The parameters of \texttt{relxillNS} model are as follows:
the inner and outer emissivity indices, $q_1$ and $q_2$,
the break radius between these two emissivity indices, $R_{\rm break}$, the dimensionless spin parameter, $a$, the inclination of the disk in units of degree, $i$, the inner and outer disk radius, $R_{\rm in}$ and $R_{\rm out}$,  in units of the innermost stable circular orbit (ISCO, $R_{\rm ISCO}$), and the gravitational
radius ($R_g = GM/c^2$), respectively, as well as the logarithm of the ionization parameter, $\log \xi$, the logarithm of the accretion disk density, $\log n$, the temperature of the input blackbody spectrum, $kT_{\rm bb}$, the iron abundance of the system normalized to the Sun,  $A_{\rm Fe}$, the reflection fraction, $f_{\rm refl}$, the redshift to the source, $z$, and the normalization of the model. This reflection model contains many parameters which are difficult to constrain simultaneously in short time intervals. We fixed the inner and outer radius of the disk at $R_{\rm ISCO}$ and $400{R_g}$, respectively. We assumed the inclination of 60$^{\circ}$ \citep[see][and references therein]{Wang17} and two emissivity indices $q_1=q_2=3$. For the NS mass, radius, and spin frequency of $1.4M_\odot$, 10 kmm and 581 Hz in 4U 1636--536, respectively, we obtained the dimensionless spin parameter $a=0.27$. We tied the temperature of the input spectrum to that of the \texttt{bbodyrad} component. We fixed the density of the accretion disk, the ionization parameter, and the iron abundance at $10^{16}~{\rm cm^{-3}}$, 3.2 and 4.5 \citep[see e.g.,][]{Wang17,Ludlam19}, respectively. The reflection fraction is fixed at -1, where the minus sign accounts for only the reflection emission. Therefore, only the normalization is free to change.  It can be seen from Figs.~\ref{fig:bb_burst1} and \ref{fig:bb_burst2} that the fitting with the disk reflection model can also achieve comparable results to fittings with the ${f\rm_a}$ model. The goodness of fit is, respectively: $\chi_{\mathrm{red}}^2/\mathrm{d.o.f.}=0.98/112$ and $\chi_{\mathrm{red}}^2/\mathrm{d.o.f.}=0.93/137$ for the two burst spectra. Then we applied the reflection model to fit the time-resolved spectra. The $\chi^2_{\rm red}$ of all burst spectra are around 1.0, which means the reflection model can also explain the deviation from the blackbody shape. We note that the reflection component contributes more than the burst emission for the first few seconds of these two PRE bursts, which has also been observed in a superburst from 4U 1636--536 \citep{Keek14} and may be attributed to the accretion disk shape \citep{He16}. 
All spectral parameters evolving over time are shown in Fig.~\ref{fig:refl}. The peak flux of two bursts are $2.3\times10^{-8}~{\rm erg~cm^{-2}~s^{-1}}$  and $3.9\times10^{-8}~{\rm erg~cm^{-2}~s^{-1}}$. Compared with the results from $f_a$ model,  the blackbody radii are very close. However, the temperatures are lower by a factor of $\sim0.25$, therefore, the burst fluxes are smaller by a factor of $\sim2.3$ for these two bursts. The evolution tracks of the blackbody radius and temperature are almost the same with the $f_a$ model.Thus, the reflection model also reveal the PRE of these two bursts.

\section{Discussion} \label{sec:highlight}


This study is based on the six type I X-ray bursts from
4U 1636--536 observed by NICER in 2017. We performed detailed time-resolved spectral analyses on two bright bursts. The persistent emission prior to the trigger of these two bursts are well fitted by \texttt{nthcomp,} and it shows a similar spectral shape. We obtained the MAXI/GSC and {\it Swift}/BAT light curves and we produced the HID from NICER observations. We found these two bursts occurred in the dip of {\it Swift}/BAT rate, with its relatively low level of hardness and high intensity implying soft spectra states. The first $\sim8$ s  of the burst spectra clearly deviated from the blackbody shape, which can be explained by the enhanced persistent emission due to Poynting-Robertson drag \citep[][]{Zand13}, as well as the reflection of the burst emission from the surrounding accretion disk \citep{Ballantyne04}. Therefore, we introduced the $f_a$ model and the reflection \texttt{relxillNS} model to fit the time-resolved burst spectra, respectively.   The $f_a$ model can fitted the burst spectra PRE, and the maximum $f_a$ are 4.7 and 6.8 for these two bursts, consistent with the values reported in \citet{Worpel13}. The peak fluxes of the bursts \#4 and \#5 are  $(4.36\pm0.62)\times10^{-8}~{\rm erg~cm^{-2}~s^{-1}}$ and  $(9.10\pm1.31)\times10^{-8}~{\rm erg~cm^{-2}~s^{-1}}$ -- slightly greater than the values powered by mixed hydrogen-helium and pure helium from the PRE bursts observed by RXTE, respectively \citep[see e.g.,][]{Galloway08}. We propose that the bursts \#4 and \#5 are powered by mixed hydrogen-helium with a solar composition and pure helium, respectively. As a comparison, the reflection model provides a comparable goodness of fit, resulting in a similar blackbody radius evolutions but smaller blackbody temperatures and thus lower burst fluxes. Since both $f_a$ and reflection models can well produce the excess below 2 keV and above 5 keV of the burst spectra, it implies that the hybrid $f_a$ and reflection models can also explain the deviations from the Planckian spectra. Moreover, it has been observed that both the Poynting-Robertson effect and the reflection from the accretion disk occurred simultaneously in a superburst from 4U 1636--536 \citep{Keek14}, and may also appear in a short X-ray burst from Aql X--1  \citep{Keek18}.  These two bursts can also contain the contributions from the enhanced persistent emission and disk reflection.  Due to lack of observations above 10 keV and the short duration of X-ray bursts, it is difficult to directly measure the proportion of these two components based solely on current observations.  If the PRE burst in ObsID 1050080128 reached the Eddington flux of helium bursts at the level of $6.4\times10^{-8}~{\rm erg~cm^{-2}~s^{-1}}$, the reflection component can contribute $\sim30\%$ of the total burst emission obtained from the $f_a$ model, similarly to the case in Aql X--1 \citep{Keek18}. For burst \#4, the accretion disk can also reflect about the same fraction of the burst emission; the  reason for this is that the persistent emission is quite similar and, hence, a comparable accretion disk morphology. For both X-ray bursts, the later stage of the burst spectra are fully consistent with the blackbody model, indicating that the cooling of the NS envelope dominates the emission without the enhanced persistent emission and reflection. Alternatively, a variable spreading layer that forms  between the NS surface and its surrounding accretion disk, can explain the change of the persistent emission \citep[see e.g.,][]{Koljonen16,Kajava17b}, where this physical picture is also applicable in 4U 1636--536. 

A few decades ago, the NS mass and radius of 4U 1636--536 were found to be in the range of  $1.28-1.65M_\odot$ and $9.1-11.3$ km from a 4.1 keV absorption line (which is unreal) and at Eddington luminosity \citep{Fujimoto86}.  \citet{Stiele16} found a radius for 4U 1636--536 that was larger than 11 km, constrained by marginally stable burning on the NS surface. Until now, the precise measurement of the NS mass and radius in 4U 1636--536 has not been achieved. The NS mass and radius in several NS LMXBs have been constrained from PRE bursts \citep[see e.g.,][]{Ozel09,Poutanen14,Li15,Li17,Suleimanov17,Suleimanov20}. The observed PRE bursts in our work have the potential to constrain the NS mass and radius of 4U 1636--536. However, the cooling tail may be polluted by the boundary layer during the soft spectral states \citep[see e.g.,][]{Poutanen14}. In practice, these PRE bursts are too short to constrain the mass and radius tightly especially for a fast-rotating NS \citep[see e.g.,][]{Suleimanov20}. In addition, the enhanced persistent emission and the reflection features are indistinguishable, making it is hard to accurately measure the Eddington flux around the touchdown moment. From the NICER and {\it NuSTAR} observations in the near future, high statistic PRE bursts in the hard spectral state, for which the Poynting-Robertson effect and disk reflection can be modelled properly, would be helpful in measuring the NS mass and radius in 4U 1636--536.

\section{Summary}
\label{sec:sum}

In this paper, we study two bright type I X-ray bursts from 4U 1636--536 in the soft spectral state based on its 2017 NICER observations.  The time-resolved spectra suggest that they belong to PRE bursts and are powered by the ignition of mixed hydrogen-helium and pure helium, respectively. We  found enhanced persistent emissions due to the Poynting-Robertson drag and disk reflections during the bursts. The contributions of the disk reflection modelled by \texttt{relxillNS} can be up to 30\% of the total burst emissions.

\begin{acknowledgements}
We thank the referee for valuable comments which improves our manuscript. We thank Javier García for generously sharing their model, \texttt{relxillNS}. ZL and YYP were supported by National Natural Science Foundation of China  (12130342, U1938107, U1838111). ZS acknowledges the supports of the National Key R\&D Program of China (2021YFA0718500) and the National Natural Science Foundation of China under grant U1838201. JL thanks supports from the National Natural Science Foundation of China under Grants Nos. 12173103, U2038101, U1938103, 11733009.  YPC thanks supports from the National Natural Science Foundation of China under Grants Nos. U1838201, U1938101. This research has made use of data obtained from the High Energy Astrophysics Science Archive Research Center (HEASARC), provided by NASA’s Goddard Space Flight Center.
\end{acknowledgements}


\bibliography{sample631}{}
\bibliographystyle{aa}
\end{document}